\documentclass[namedreferences]{solarphysics}
\usepackage[optionalrh]{spr-sola-addons}
\usepackage{hyperref}
\usepackage{graphicx}
\usepackage{url}

\begin{document}

\begin{opening}

\title{The Relationship between Solar Coronal X-Ray Brightness and Active Region Magnetic Fields: A Study Using High-Resolution {\it Hinode} Observations}
\author{Soumitra \surname{Hazra}$^{1,2}$ \sep
        Dibyendu \surname{Nandy}$^{1,2}$ \sep \\
        B \surname{Ravindra}$^{3}$
       }
\runningtitle{Relationship between Coronal X-Ray Brightness and Active Region Magnetic Fields}
\runningauthor{S. Hazra \it {et al.}}
\institute{$^{1}$~Department~of~Physical~Sciences,~Indian~Institute~of~Science~Education~and~Research,~Kolkata, Mohanpur~741246, West Bengal, India
              $^{2}$~Center~of~Excellence~in~Space~Sciences~India,~IISER~Kolkata,~Mohanpur~741246,~West~Bengal,~India \\
                     email: \url{s.hazra@iiserkol.ac.in} \\
                     email: \url{dnandi@iiserkol.ac.in} \\
              $^{3}$~Indian~Institute~of~Astrophysics,~Koramangala,~Bengaluru~560034,~India
                     email: \url{ravindra@iiap.res.in} \\
              }

\begin{abstract}
By using high-resolution observations of nearly co-temporal and co-spatial {\it Solar Optical Telescope} spectropolarimeter
and {\it X-Ray Telescope} coronal X-ray data onboard {\it Hinode}, we revisit the problematic relationship between global magnetic quantities and coronal X-ray brightness. Co-aligned vector magnetogram and X-ray data were used for this study. The total X-ray brightness over active regions is well correlated with integrated magnetic quantities such as the total unsigned magnetic flux, the total unsigned vertical current and the area-integrated square of the vertical and horizontal magnetic fields. On accounting for the inter-dependence of the magnetic quantities, we inferred that the total magnetic flux is the primary determinant of the observed integrated X-ray brightness. Our observations indicate that a stronger coronal X-ray flux is not related to a higher non-potentiality of active-region magnetic fields. The data even suggest a slightly negative correlation between X-ray brightness and a proxy of active-region non-potentiality. Although there are small numerical differences in the established correlations, the main conclusions are qualitatively consistent over two different X-ray filters, the Al-poly and Ti-Poly filters, which confirms the strength of our conclusions and validate and extend earlier studies that used low-resolution data. We discuss the implications of our results and the constraints they set on theories of solar coronal heating.
\end{abstract}

\end{opening}

\section{Introduction}
Solar active-region coronal loops appear bright in EUV and X-ray wavelengths, which is indicative of
very high temperatures of the order of a million degrees Kelvin. The origins
behind these high-temperature coronal structures remain elusive. An energy flux of about 
10$^{7}$~ergs~cm$^{2}$~s$^{-1}$ is required to maintain this high
temperature of the coronal plasma \cite{withbr77}. It has been suggested
that there is a one-to-one correspondence between the location of the magnetic fields
in the photosphere and bright coronal structures in the corona \cite{vaia73} and we know that
most of the coronal X-ray luminosity is concentrated within active-region magnetic-flux systems.

Several theories have been proposed to explain the heating of coronal structures \cite{zirk93, nar96,
ascn04, klim06}. These theories are broadly classified into two
subcategories: DC heating model, {\it i.e.} the nano-flare heating model \cite{park88}, and
the AC heating model, {\it i.e.} the wave-heating theory ({\it e.g.} see the review by \opencite{ascn04}).
In the AC heating model, high-frequency MHD waves are generated in the magnetic foot points
of active regions and propagate through magnetic loops in the corona. These
waves dissipate their energy in the corona \cite{nar96}. Although recent observations reveal
that MHD waves propagate into the quiet solar corona \cite{tomc07}, it is unclear
whether these MHD waves alone can heat the corona to such
a high temperature \cite{mandr20, cirt13}. Alternatively, DC-heating models are proposed
to explain the heating of active regions where nanoflare-like small bursts
(each of energy $10^{24}$ erg) can liberate energy by magnetic reconnection -- driven by the constant shuffling of magnetic foot points by turbulent convective motions just beneath the photosphere \cite{park88, cirt13}. It has been suggested recently that waves can play a major role in heating the quiet-Sun corona \cite{mcint11, wedem12}, while for coronal active regions the additional DC heating mechanism must play a role \cite{park88, klim06}.

To examine the relative roles of diverse physical mechanisms in the context of coronal heating,
it is essential to have information on the coronal magnetic and velocity field. Current instrumentation is still at a nascent stage, however, and is inadequate for such coronal diagnostics \cite{lin04}. Since coronal field lines are linked to the photosphere, another approach is possible: exploring the relationship between photospheric
magnetic-field parameters and brightness of the coronal loops.
In earlier studies, \inlinecite{fish98} and \inlinecite{tan07} investigated the
relationship between the X-ray luminosity and photospheric magnetic-field parameters.
These two studies reported a strong correlation between the X-ray luminosity and the total unsigned
magnetic flux. \inlinecite{tan07} also found a good correlation between the
average X-ray brightness and average Poynting flux, but ruled out any correlation between the velocity of footpoint motions and total X-ray brightness. Their computed Poynting flux had a range between 10$^{6.7}$ and
10$^{7.6}$~ergs~cm$^{-2}$~s$^{-1}$, which is enough to heat the corona \cite{withbr77}.
Using data from other wavelengths (UV/EUV channels), \inlinecite{chand13} also found
a good correlation between total emission from bright points and total unsigned photospheric magnetic flux.
Forward-modelling of active regions also suggests a direct correlation between magnetic flux and X-ray luminosity \cite{lund08}.

The net current is a measure of non-potentiality of the magnetic field in the active region.
In active-region flare and coronal-mass-ejection processes, the non-potentiality of the magnetic field
can play a significant role \cite{sch06,jing06, wan08}. As a result of the low resistivity, large-scale ($10^3$ km)
currents cannot dissipate sufficiently in the corona \cite{hagya88}; thus these currents may have no contribution to
coronal heating.  Earlier observations do not find a strong relationship between the total X-ray luminosity and total vertical current \cite{metc94b, fish98}. Note that \inlinecite{wan08} showed the existence of 3D current structures over active regions. Another traditionally used measure for magnetic non-potentiality is the parameter $\alpha_\textrm{best}$ which appears in the force-free field equation and which is thought to be related to the wrapping of magnetic-field lines along the axis of an active-region flux tube ({\it i.e.} the twist of magnetic-field lines). Observation shows that there is no significant correlation between X-ray brightness and $\alpha_\textrm{best}$ \cite{fish98,nandy08}. While many studies have used $\alpha_\textrm{best}$ as the measure of the twist in the solar active region, this is questionable because the photosphere is not deemed to be force-free \cite{leka05}. Other studies have shown that a polarity-inversion line near coronal-loop foot points and strong magnetic shear may also result in enhanced coronal emission (\opencite{falc97a}; \opencite{falc97}, \citeyear{falc20}).

\inlinecite{long96} proposed the minimum current corona (MCC) model where coronal heating was described as a
series of small reconnection events punctuating the quasi-static evolution of coronal field.
This model qualitatively predicts the variation of the X-ray luminosity with the total flux that closely
matches observations \cite{fish98}. \inlinecite{wan20} have observed bright coronal loops and
diffused coronal loops that are associated with the quasi-separatrix layers (QSLs). Since QSLs
are the places where energy release occurs through 3D magnetic reconnection, they
concluded that QSLs are important for heating the active-region corona and chromosphere.
By analysing the X-ray images taken from {\it Hinode/X-Ray Telescope} (XRT) and corresponding {\it Michelson Doppler Imager} (MDI) line-of-sight
magnetograms, \inlinecite{lee10} found a relationship between coronal-loop brightness and
magnetic topologies in AR 10963. They also found that frequent transient brightenings in coronal
loops are related to separators that have a large amount of free energy.

Here we revisit the coronal-heating problem with space-based vector-magnetogram data, which are free from atmospheric seeing effects, which can produce cross talk between various Stokes parameters. Such space-based magnetic field measurements have also reduced atmospheric scattered light contribution. The obtained vector-field data are of very high resolution, thereby reducing the effect of filling factor. In this article, we use X-ray images taken from two filters (Ti-poly and thin Al-poly) of the XRT telescope onboard the {\it Hinode} spacecraft and vector magnetic-field measurements taken from the {\it Spectro-Polarimeter} (SP) of {\it Solar Optical Telescope} (SOT) to study the relationship between the X-ray brightness and magnetic-field parameters in active-region flux systems. This study extends previous work that used lower resolution \textit{Yohkoh} data \cite{fish98}. We also, incidentally, explore the effect of the filter response (which is mainly affected by deposition of unknown materials on CCD cameras) on the relationship between X-ray brightness and magnetic-field parameters. In Section 2 we provide the details of the data used in this study. In Section 3, we detail our results. In Section 4, we discuss the implication of our results for the heating of the solar corona.

\section{Data Analysis}
\subsection{Data Selection}
The {\it X-ray telescope} (XRT: \opencite{golu07}) onboard the {\it Hinode} spacecraft \cite{kosu07}
takes images of the solar corona at a spatial resolution of one arcsec per pixel using different filters.
XRT images are of the size 2k$\times$2k pixel, which covers a 34$\times$34 square arcmin
field of view (FOV) of the solar corona. XRT observes coronal
plasma emission in the temperature range $5.5 <\textrm{log} T < 8$, which is realized by different X-ray
filters, that have their own passband, corresponding to different responses to plasma temperature.
Within a few months of the launch of the {\it Hinode} spacecraft, contaminating materials were deposited on the CCD,
which significantly impacted the filter response, specifically for observations of longer
wavelengths. Regular CCD bakeouts were unable to completly remove this contamination.
As the effect of the contamination is mainly wavelength-dependent (the long-wavelength observations are affected more strongly), the observations from the thin Al-poly/Al-mesh
filter are more heavily affected than the other filters such as Ti-poly, and Be-med. For the
present study, we have used data taken from Ti-poly and the thin Al-poly filter,
which observe the solar coronal plasma at temperatures higher than 2 MK and 0.5 MK,
respectively. Therefore, we have a point of comparison to establish whether filter degradation may play a role in
the inconsistencies of the results.\\
\~~~~~The {\it Spectro-Polarimeter} (SP: \opencite{ichim08}) is a separate back-end instrument of
the {\it Solar Optical Telescope} (SOT: \opencite{tsun08}) onboard the {\it Hinode} spacecraft. The SP provides 
Stokes signals with high polarimetric accuracy in the 6301 and 6302~\AA~photospheric
lines. The primary product of the Stokes polarimeter are the Stokes-{\it IQUV} profiles, which are suitable for deriving the vector
magnetic field in the photosphere. The spatial resolution along the slit direction is
0.295$^{\prime\prime}$~pixel$^{-1}$; in the scanning direction it is 0.317$^{\prime\prime}$~pixel$^{-1}$.
The Stokes vector was inverted using the MERLIN code, which is based on the Milne--Eddington
inversion method. The inverted data provide the field strength, inclination and azimuth
along with the Doppler velocity, continuum images, and many other parameters. The processed
data were obtained from the Community Spectropolarimetric Analysis Center (\href{http://www.csac.hao.ucar.edu/}{CSAC}).
We corrected for the ambiguity in the transverse component of the magnetic field using the minimum-energy algorithm \cite{metc94, Leka2009}. The resulting
magnetic-field vectors were transformed into heliographic co-ordinates
\cite{venkat89}. We selected 40 different NOAA active regions observed at different times of the year. We also excluded active regions whose central meridional
distance was greater than $30^{\circ}$. We took the vector magnetogram data close to the
timings of soft X-ray data obtained from both the Ti-poly and the Al-poly filter of the XRT.
In Table 1 and 2 we list the different active regions used in this study, the date and time of
the observations of the vector magnetogram, and the corresponding soft X-ray data. After these two sets of data, we also obtained the G-band data taken by the {\it X-Ray Telescope}. These data were used for to co-align each of the data sets. For each selected vector magnetogram, we simulteniously took
XRT X-ray (Ti-poly and Al-poly) data and G-band data. Throughout this article,
we use the term Ti-poly dataset to represent the X-ray image obtained from the Ti-poly filter of
XRT onboard {\it Hinode}.  Similarly, we use the term Al-poly for the data taken from the Al-poly filter.
There is always a corresponding vector magnetogram associated with these data sets.
The X-ray data were calibrated using the xrt$\_$prep.pro available in the Solarsoft
routines. The calibrated data were normalized to a one-second exposure time.
 \begin{table}[]
\caption{NOAA active regions and time of corresponding XRT X-ray Ti-poly filter and SP magnetogram data}
\centering
\begin{tabular}{r  c  c  c}
Date & NOAA  & Magnetogram scan &  XRT X-ray (Ti-Poly)\\
     & active region & start time [UT]& observation time [UT]\\ \hline
    1 May 2007 & 10953 & 05:00:04 & 05:00:57  \\ 
    1 Jul. 2007 & 10962 & 13:32:05 & 13:31:51 \\ 
   15 Jul. 2010 & 11087 & 16:31:19 & 16:30:53 \\ 
    10 Aug. 2010 & 11093 & 09:15:04 & 09:14:18  \\ 
    31 Aug. 2010 & 11102 & 02:30:04 & 02:30:42  \\ 
    23 Sep. 2010 & 11108 & 07:21:05 & 07:21:12  \\ 
    26 Oct. 2010 & 11117 & 10:45:46 & 10:51:55  \\ 
    22 Jan. 2011 & 11149 & 09:31:28 & 09:43:22  \\ 
    14 Feb. 2011 & 11158 & 06:30:04 & 06:30:02  \\ 
    4 Mar. 2011 & 11164 & 06:15:06 & 06:15:04  \\ 
    31 Jan. 2012 & 11411 & 04:56:32 & 04:57:24  \\ 
    18 Feb. 2012 & 11419 & 11:08:53 & 11:10:10  \\ 
    8 Mar. 2012 & 11429 & 21:30:05 & 21:32:22 \\ 
    22 Apr. 2012 & 11463 & 04:43:05 & 04:48:31  \\ 
    12 May 2012 & 11476 & 12:30:50 & 12:30:41  \\ 
    18 May 2012 & 11479 & 04:47:05 & 04:48:38 \\ 
    5 Jul. 2012 & 11517 & 03:45:35 & 04:18:11  \\ 
   12 Jul. 2012  & 11520 & 11:12:28 & 11:12:45  \\ 
   14 Aug. 2012  & 11543 & 14:35:05 & 14:35:30 \\ 
   25 Sep. 2012  & 11575 & 12:49:06 & 12:50:08 \\ 
    \end{tabular}
\end{table}

\begin{table}[]
\caption{NOAA active regions and time of corresponding XRT X-ray Al-poly filter and SP magnetogram data}
\centering
\begin{tabular}{r  c  c  c}
 Date & NOAA & Magnetogram scan  &  XRT X-ray (Al-Poly) \\
     & active region & start time [UT]& observation time [UT]\\ \hline
    30 Aug. 2011 & 11280 & 07:35:23 & 07:35:36  \\ 
    13 Sep. 2011 & 11289 & 10:34:05 & 10:34:24 \\ 
    28 Sep. 2011 & 11302 & 18:38:05 & 18:38:16 \\ 
    28 Nov. 2011 & 11360 & 00:05:20 & 00:03:04  \\
    31 Jan. 2012 & 11410 & 04:56:32 & 05:25:10  \\ 
    1 Feb. 2012 & 11413 & 08:51:31 & 09:03:35  \\ 
    8 Mar. 2012 & 11429 & 01:20:05 & 01:23:50 \\
    22 Apr. 2012 & 11463 & 04:43:05 & 04:55:22  \\
    16 Aug. 2012 & 11543 & 13:35:05 & 13:35:19  \\ 
    25 Sep. 2012 & 11575 & 12:49:06 & 12:50:37  \\ 
    2 Oct. 2012 & 11582 & 09:53:06 & 09:54:39  \\ 
    17 Oct. 2012 & 11589 & 09:06:01 & 09:06:22  \\
    28 Oct. 2012 & 11594 & 01:40:05 & 01:42:23  \\ 
    17 Nov. 2012 & 11613 & 10:25:06 & 10:25:37 \\
    17 Nov. 2012 & 11619 & 12:49:06 & 12:50:37  \\ 
    10 Feb. 2013 & 11667 & 14:30:04 & 14:31:22  \\ 
    15 Mar. 2013 & 11695 & 09:30:51 & 09:33:22 \\ 
    31 Aug. 2013 & 11836 & 18:14:36 & 18:15:25 \\ 
    27 Sep. 2013 & 11850 & 09:30:05 & 09:30:06 \\ 
    \end{tabular}
\end{table}

\par
\subsection{Data Coalignment}
 To overlay the XRT X-ray data with vector magnetograms, we first co-aligned the G-band
data taken by XRT telescope with the continuum image. The continuum image was obtained by inverting the Stokes data set. 
To do this, we first identified the dark center of the
sunspot in the G-band and continuum images. Later, we interpolated the continuum image data
to the XRT image resolution. In the next step, we choose the same field of view (FOV) in the two data sets. 
By using the maximum-correlation method, we then co-aligned the continuum images
with the G-band images. A similar shift was applied to the vector field data to
co-align the entire dataset with X-ray images of the XRT dataset.

\section{Integrated Quantities}
We derived various integrated quantities and compared them with the X-ray
brightness. We computed the individual as well as integrated quantities such as total
magnetic flux, and total magnetic energy {\it etc} and compared them with the
X-ray brightness. Below, we describe each of these quantities.

\subsection{Active-Region Coronal X-Ray Brightness}
The integrated X-ray brightness [Lx] was computed by summing the values of each
bright pixel in the image and then multiplying by the pixel area. The bright pixels were
selected by using the threshold values. We found the rms value in the X-ray image and selected
only those pixels whose value were higher than the 1-$\sigma$ level (the rms value) of the image.

\subsection{Global Magnetic-Field Quantities}
Since our selected active regions are close to the disk centre, the magnetic-field
vectors are horizontal and vertical to the solar surface.  Using the $B_{x}$, $B_{y}$, and
 $B_{z}$ components, it is possible to define the integrated quantities, which can be correlated with the
X-ray brightness to find the relationship between the two (for detailed information about
integrated quantities, see \opencite{fish98}, \opencite{leka07}). We selected pixels in
$B_{x}$, $B_{y}$ and $B_{z}$ whose values are greater than the 1-$\sigma$ level of these images. The
following integrated quantities were computed from magnetic field components:

\begin{equation}
\phi_\textrm{tot}=\sum |B_{z}| \textrm{d} A
\end{equation}
\begin{equation}
B_{z,\textrm {tot}}^2=\sum B^{2}_{z}\textrm{d} A
\end{equation}
\begin{equation}
B^2_{h,\textrm{tot}}=\sum B^2_{h}\textrm{d} A
\end{equation}
\begin{equation}
J_\textrm{tot}=\sum |J_{z}|\textrm{d} A
\end{equation}

Here $B_{z}$ and $B_{h}$ represent the vertical and horizontal magnetic field,  $J_{z}$ is the vertical
current density, and $\displaystyle\sum \textrm{d}A$ is the effective area on the solar surface. Since the ratio of the vertical current density and magnetic field is related to the handedness or chirality (twist) of the underlying flux tube \cite{long98}, we also introduced a quantity $\mu_0 J_{\textrm{tot}}/\phi_{\textrm{tot}}$ (ratio of unsigned total current and unsigned total magnetic flux), which has the same units as the twist and can thus be taken as a proxy for it.  Highly twisted flux tubes are strongly non-potential, and thus the quantity above is a measure of the non-potentiality of active region flux systems.

We computed all of the magnetic quantities from the vector magnetogram for all active regions. The average estimated errors of the magnetic variables: $B_z$, $J_z$, $B^2_z$, and $B^2_h$ are 8 G, 45 mA, 64 G$^2$, and 800 G$^2$.

\begin{figure}[!h]
\begin{center}
\includegraphics[angle=90,width=1.00\textwidth]{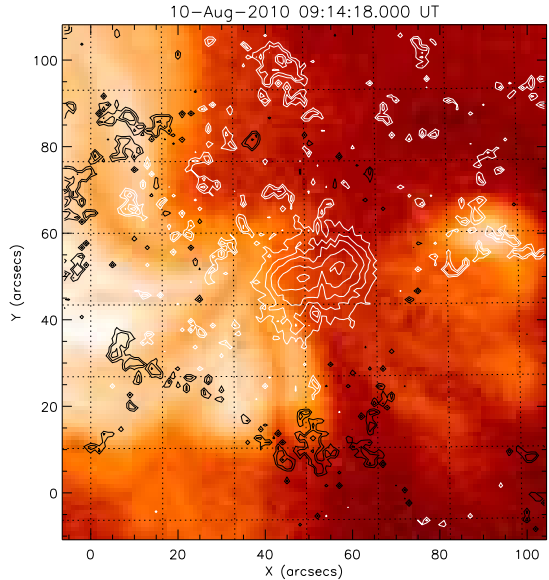}
\end{center}
\caption{\footnotesize The contours of vertical magnetic field overlaid upon the X-ray
image of Active Region NOAA 11093 taken in Ti-poly filter by the XRT telescope. Contours with thick solid lines (white) represent the
positive magnetic fields with a field strength level of 500, 1000, 1500, 2000, and 3000 G; thin solid
lines (black) represent the negative vertical magnetic field at the same level.}
\label{fig:1}
\end{figure}
\begin{figure*}[!htb]
 \begin{center}
 \centering
\Large\Huge\includegraphics[angle=90,width=1.00\textwidth]{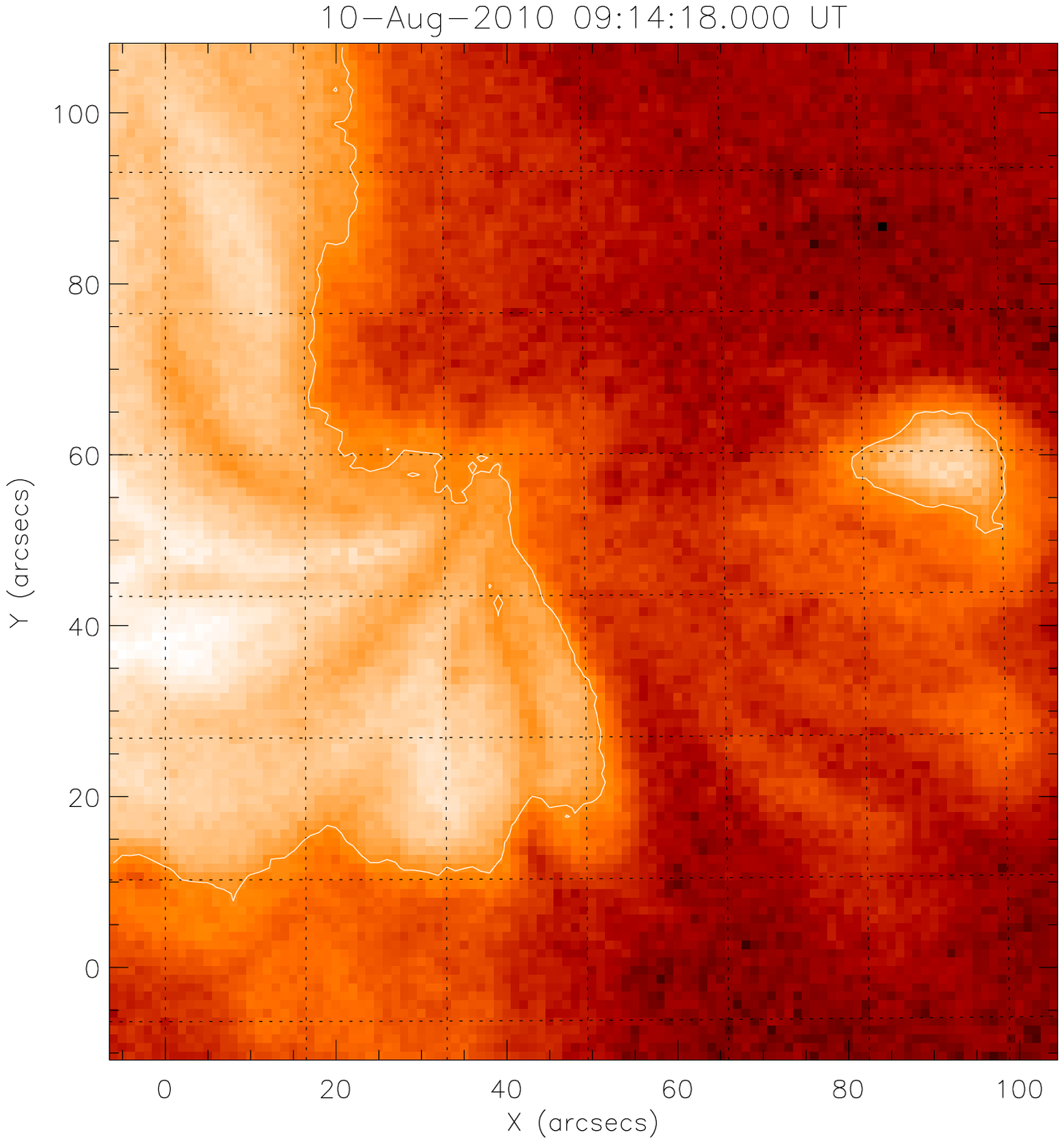}
\end{center}
\caption{\footnotesize Contour map of the 1$\sigma$ level of X-ray brightness
overlaid on the X-ray image of the Active Region NOAA 11093. }
 \label{fig:2}
 \end{figure*}

\section{Results}
Figure \ref{fig:1} shows the contours of the $B_{z}$-component of the magnetic
field overlaid on the X-ray image of active region NOAA 11093 after co-aligning the images. 
The contour map shows that the X-ray brightness in the corona overlying the umbral part of the
sunspot is lower than that of the loops emanating from the penumbral part of the active region. The bright loops
are associated with the plage regions as has been observed before \cite{palla79}.
The loops are still not fully resolved in the XRT images, but the
cluster of loops clearly turn in a clockwise direction. On the west side of the sunspot, the loop
structures are absent. At the same location in the photosphere, large-scale plage structures
are also absent. This may indicate that large-scale plage regions are essential for
the loops to appear in X-rays. Thus we note that a visual spatial correlation exists between the
location of the plages and the bright loops in X-rays.\\

\subsection{Correlation Between Global Magnetic Field Quantities and X-ray Brightness}

We explored the relationship between total (area-integrated) magnetic quantities and X-ray brightness in active regions.
We used the XRT data for 20 active regions each in the Ti-poly and
Al-poly data sets (all data are listed in Tables 1 and 2). We
only selected those pixels whose intensity values exceeded a 1-$\sigma$ threshold in the
X-ray and magnetic images. Figure~\ref{fig:2} shows the contour map of 1-$\sigma$ level threshold of X-ray brightness
overlaid upon the X-ray image of Active Region NOAA 11093. The 1-$\sigma$ level threshold line of the contour map clearly indicates the borders of the bright loops.

\begin{figure}[!h]
 \begin{center}
\includegraphics[width=1.0\textwidth]{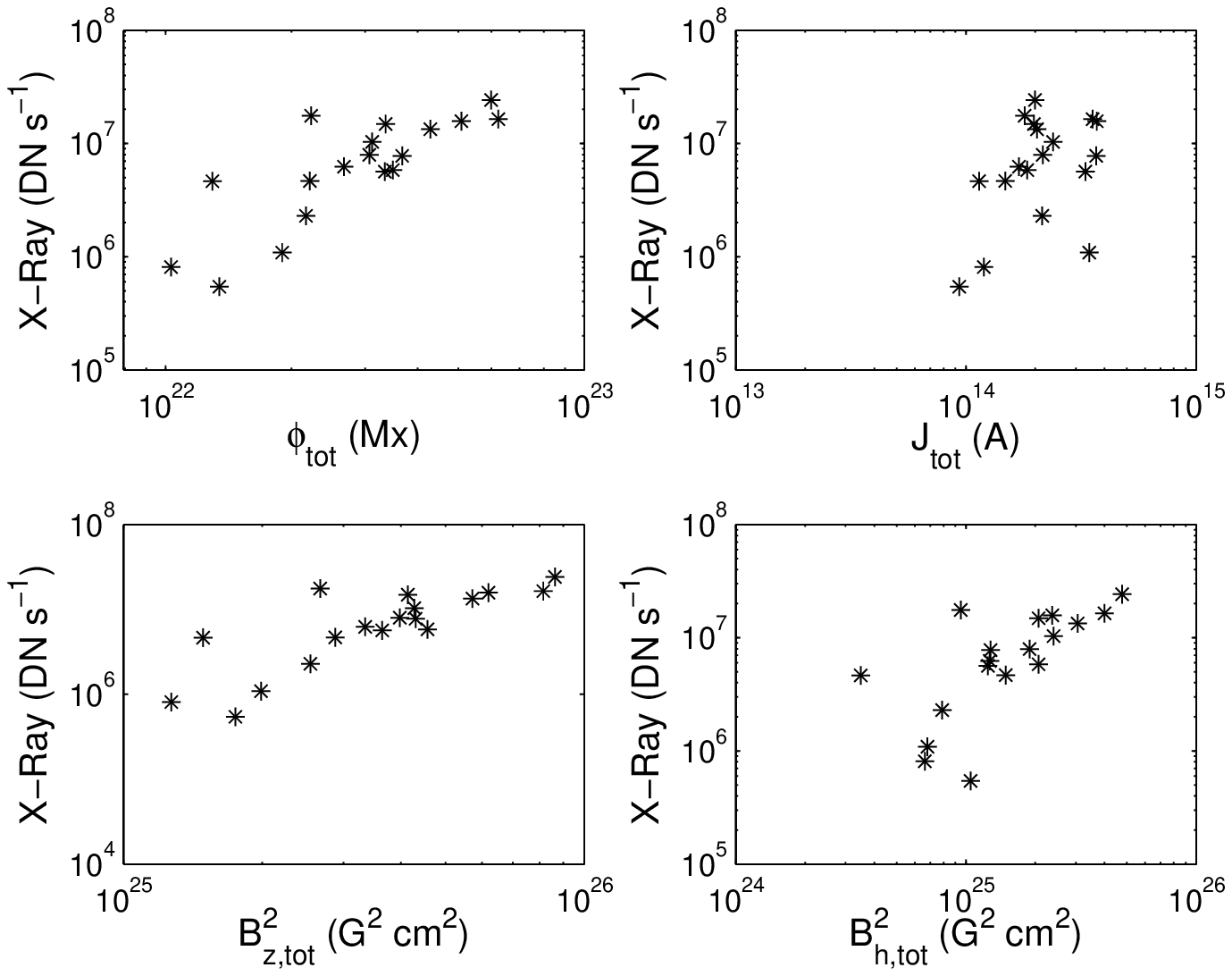}
\end{center}
\caption{\footnotesize Relationship between X-ray brightness and global magnetic-field quantities $\phi_{\textrm{tot}}$, $J_{\textrm{tot}}$, $ B^2_{z,\textrm{tot}}$, and $ B^2_{h,\textrm{tot}}$  (using the data set of Table 1, {\it i.e.} the Ti-poly filter). Correlation coefficients are listed in Table 3.}
\label{fig:3}
 \end{figure}
 \begin{figure}[!h]
 \begin{center}
\includegraphics[width=1.0\textwidth]{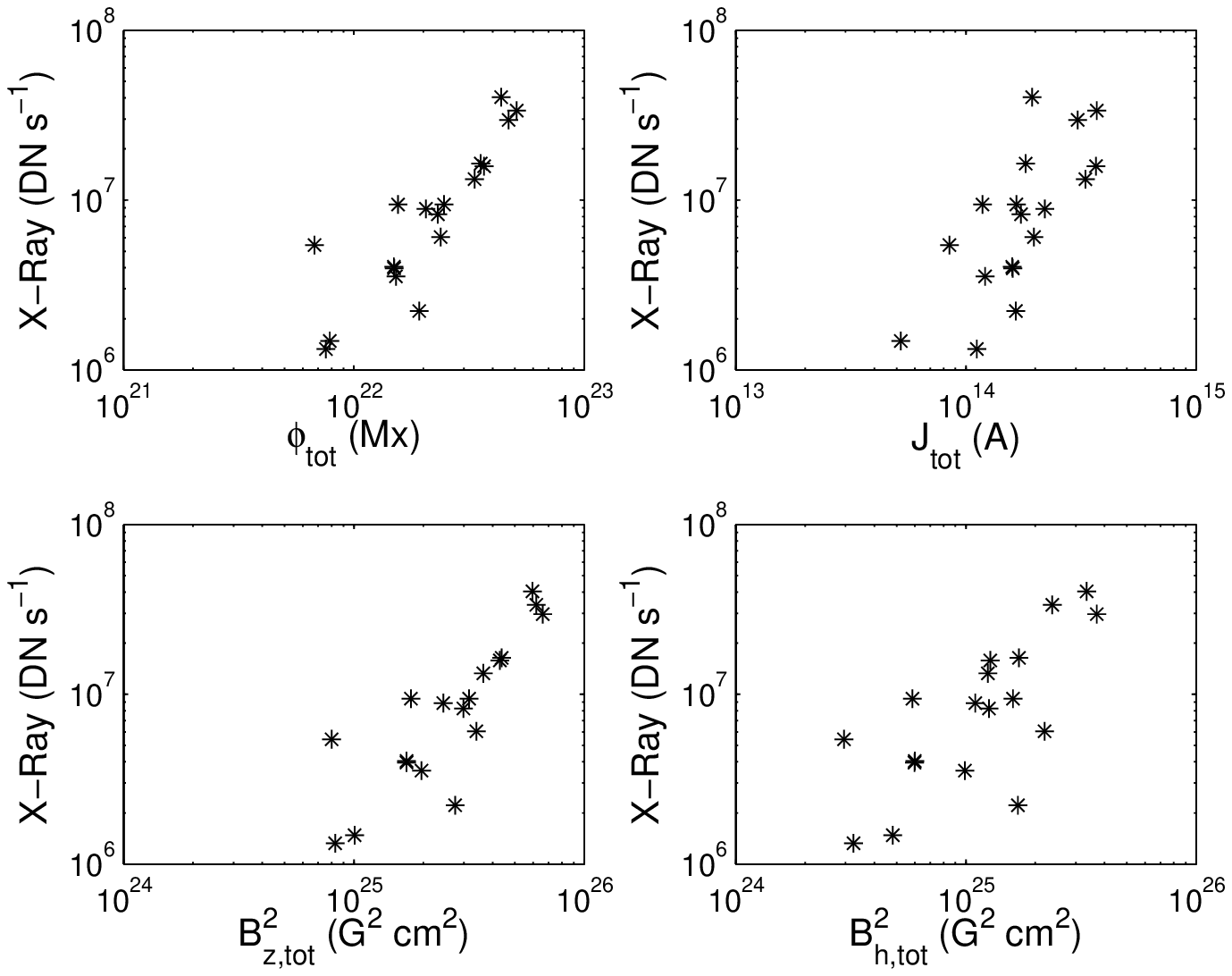}
\end{center}
\caption{\footnotesize Relationship between X-ray brightness and global magnetic-field quantities $\phi_{\textrm{tot}}$, $J_{\textrm{tot}}$, $ B^2_{z,\textrm{tot}}$, and $ B^2_{h,\textrm{tot}}$  (using the data set of Table 2, {\it i.e.} the Al-poly filter). Correlation coefficients are listed in Table 3.}
\label{fig:4}
\end{figure}

Figures~\ref{fig:3} and \ref{fig:4} depict the relationship between the X-ray brightness and total unsigned magnetic flux (top-left), $B^{2}_{z,\textrm{tot}}$ (top-right), $B^{2}_{h,\textrm{tot}}$ (bottom-left), and unsigned $J_{\textrm{tot}}$ (bottom-right)
in logarithmic scale.  Figure~\ref{fig:3} is for the data sets of Table 1, {\it i.e.} Ti-poly filter and Figure~\ref{fig:4}
is for the data sets of Table 2, {\it i.e.} the Al-poly filter. The coronal X-ray brightness and
the global magnetic-field parameters in both data sets are clearly correlated; although the correlation coefficients are
numerically somewhat different, they are qualitatively similar (for quantative correlation coefficients see Table 3).

\begin{figure}[!h]
 \begin{center}
\includegraphics[width=1.0\textwidth]{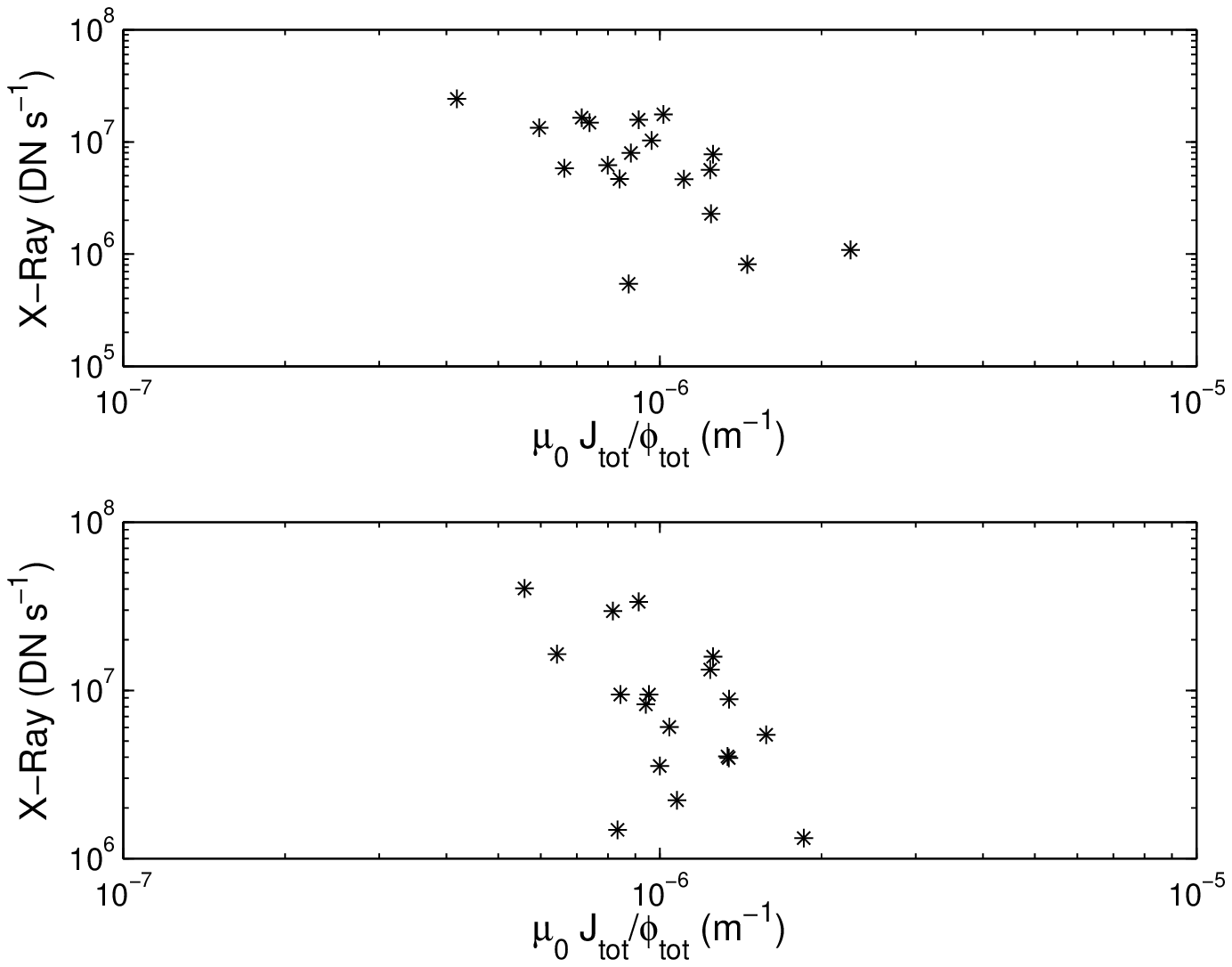}
\end{center}
\caption{\footnotesize Scatter plots of X-ray brightness with $\mu_0 J_{tot}/\phi_{tot}$ (top plot is for the data set of Table 1, {\it i.e.} the Ti-poly filter, the bottom plot is for the data set of Table 2, {\it i.e.} the Al-poly filter). Correlation coefficients are listed in Table 3}
\label{fig:5}
 \end{figure}

Non-potential flux systems are known to be storehouses of free energy, and it is often assumed that therefore, a coronal energy release in X-rays should be positively correlated with measures of non-potentiality. Figure \ref{fig:5} depicts the relationship between X-ray brightness and the non-potentiality measure $\mu_{0}J_{\textrm{tot}}/\phi_{\textrm{tot}}$. The top plots are for data sets of Table (Ti-poly filter data), the bottom plots are for data sets of Table 2 (Al-poly filter data). The X-ray brightness is anti-correlated with $\mu_{0}J_{\textrm{tot}}/\phi_{\textrm{tot}}$ in both cases.

To determine which of the magnetic quantities contributes predominantly to the X-ray brightness, we need to  examine
whether there is any inter-dependence between the global magnetic quantities. In Section 4.2, we follow \inlinecite{fish98} in this analysis and establish the correlation between each of the magnetic parameters with the total unsigned flux first and also perform a partial correlation analysis to extract the true underlying dependencies.

\begin{table}[!h]
\caption{Correlation coefficients between different parameters}
\centering
\begin{tabular}{c c c c}
Figure &  Correlated quantities  &   Pearson correlation & Spearman correlation  \\
 number    &   & coefficients with & coefficients with\\
     &   & confidence levels &  confidence levels\\

\hline
Figure 3 & X-ray brightness {\it vs.} $\phi_{\textrm{tot}}$ & 0.81 ($99.99 \%$) & 0.83 ($99.99 \%$)\\
(Ti-Poly)       & X-ray brightness {\it vs.} $J_{\textrm{tot}}$ & 0.31 ($98.52 \%$) & 0.50 ($96.63 \%$)\\
       & X-ray brightness {\it vs.} $B^{2}_{z,\textrm{tot}}$ & 0.81 ($99.99 \%$) & 0.75 ($99.99 \%$)\\
       & X-ray brightness {\it vs.} $B^{2}_{h,\textrm{tot}}$ & 0.79 ($99.98 \%$) & 0.79 ($99.98 \%$)\\ \hline
Figure 4 & X-ray brightness {\it vs.} $\phi_{\textrm{tot}}$ & 0.90 ($99.99 \%$) & 0.89 ($100 \%$)\\
(Al-Poly)       & X-ray brightness {\it vs.} $J_{\textrm{tot}}$ & 0.62 ($99.99 \%$) & 0.76 ($100 \%$)\\
       & X-ray brightness {\it vs.} $B^{2}_{z,\textrm{tot}}$ & 0.91 ($99.99 \%$) & 0.71 ($99.99 \%$)\\
       & X-ray brightness {\it vs.} $B^{2}_{h,\textrm{tot}}$ & 0.81 ($99.67 \%$) & 0.85 ($99.99 \%$)\\ \hline

Figure 5 & X-ray brightness {\it vs.}  & -0.59 ($96.91 \%$) & -0.54 ($82.16 \%$)\\
       & $\mu_0 J_{\textrm{tot}}/\phi_{\textrm{tot}}$ (Ti-poly)& &\\
       & X-ray brightness {\it vs.}  & -0.55 ($97.1 \%$) & -0.54 ($97.92 \%$)\\
       & $\mu_0 J_{\textrm{tot}}/\phi_{\textrm{tot}}$ (Al-poly)& &\\ \hline

Figure 6 & $J_{\textrm{tot}}$ {\it vs.} $\phi_{\textrm{tot}}$ & 0.57 ($99.99 \%$) & 0.63 ($100 \%$)\\
       & $B^{2}_{z,\textrm{tot}}$ {\it vs.} $\phi_{\textrm{tot}}$ & 0.99 ($99.99 \%$) & 0.98 ($99.99 \%$)\\
       & $B^{2}_{h,\textrm{tot}}$ {\it vs.} $\phi_{\textrm{tot}}$ & 0.91 ($99.99 \%$) & 0.88 ($99.99 \%$)\\
       & $\mu_0 J_{\textrm{tot}}/\phi_{\textrm{tot}}$ {\it vs.} $\phi_{\textrm{tot}}$ & -0.54 ($98.86 \%$) & -0.61 ($99.94 \%$)\\
\end{tabular}
\label{tab:3}
\end{table}

\subsection{Correlations Among Global Magnetic-Field Quantities and Partial Correlation Analysis}
\begin{table}[!h]
\caption{Partial correlation coefficients between different quantities for different filters}
\centering
\begin{tabular}{c c c } 
\hline
Correlated quantities&\multicolumn{2}{c}{Partial Correlation Coefficient} \\
 (controlling $\phi_{\textrm{tot}}$ ) &  Ti-Poly  &   Al-Poly \\
          &  (Table 1 data set) & (Table 2 data set)\\
\hline
X-ray brightness {\it vs.} $J_{\textrm{tot}}$   & -0.45  & -0.64 \\ 
X-ray brightness {\it vs.} B$^{2}_{z,\textrm{tot}}$   & 0.37  & 0.29\\ 
X-ray brightness {\it vs.} B$^{2}_{h,\textrm{tot}}$   & 0.32  & 0.25\\ 
\end{tabular}
\label{tab:4}
\end{table}

\begin{figure}[!h]
 \begin{center}
\includegraphics[width=1.0\textwidth]{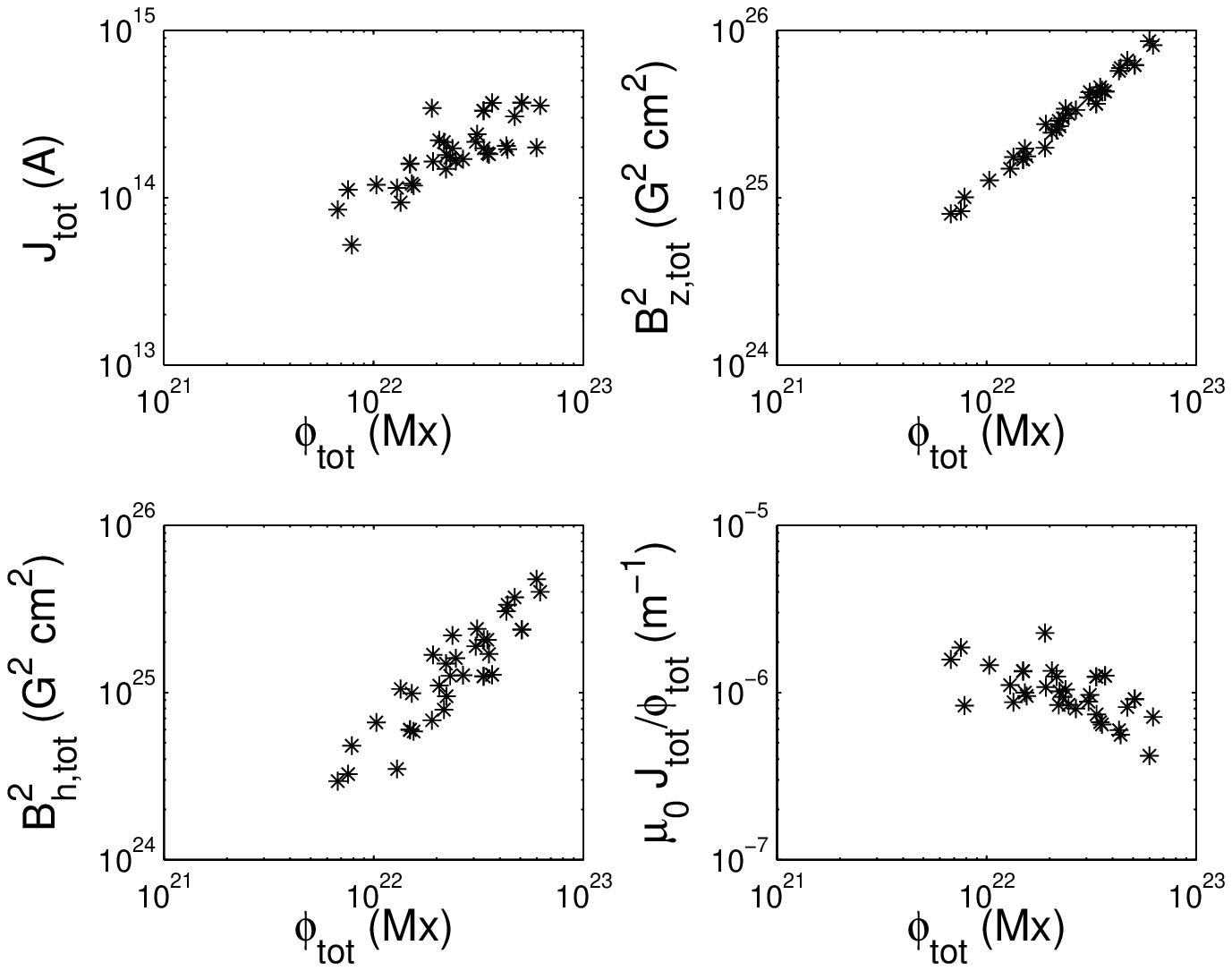}
\end{center}
\caption{\footnotesize Relationship of global magnetic quantities $J_{\textrm{tot}}$, $B^2_{z,\textrm{tot}}$, $B^2_{h,\textrm{tot}}$, and $\mu_0 J_{\textrm{tot}}/\phi_{\textrm{tot}}$ with $\phi_{\textrm{tot}}$. Correlation coefficients are listed in Table 3.}
\label{fig:6}
\end{figure}

Figure~\ref{fig:6} (see also Table 3) shows the inter-dependence of the total unsigned magnetic flux and
all other magnetic variables, such as the total absolute current, $B^{2}_{z,\textrm{tot}}$, $B^{2}_{h,\textrm{tot}}$, and
$\mu_0 J_{\textrm{tot}}/\phi_{\textrm{tot}}$.  Each of these magnetic parameters shows a good correlation with
the total unsigned magnetic flux, which means that they are related to each other through size (area integration).
To find the relationships between the different magnetic parameters, we carried out a partial-correlation analysis. In the partial-correlation technique, the correlation between the two dependent variables is
examined after removing the effects of other variables.
Table~\ref{tab:4} shows the partial correlation coefficients between the X-ray
brightness and integrated magnetic quantities (except for the magnetic flux) after removing the
effect of magnetic flux. Again, we find that although the correlation coefficients are numerically somewhat different, they are qualitatively similar across the two filters. We do not find any significant correlation between X-ray brightness and other magnetic quantities (except for a slightly negative correlation for $J_{\textrm{tot}}$, which is lower for the Ti-poly filter). Thus, it appears that the total magnetic flux is the primary positive contributor to the total coronal X-ray flux over solar active regions.

\subsection{Filter Issues in the X-ray Data}

Our analysis shows that there are minor differences in the established relationships gleaned from the Ti-poly and Al-poly data sets. We suggest that this small difference in results can be explained as a consequence of contamination in CCDs that could have altered the filter response. The Ti-poly X-ray data and Al-poly X-ray data have strong linear correlation (linear correlation coefficient 0.99) which indicates that there are no calibration problems with the XRT data (see Figure 7). Taken together with the fact that the results are qualitatively similar from both filters, this lends strong credence to the data and our conclusions.

\begin{figure}[!h]
 \begin{center}
\includegraphics[angle=90, width=1.0\textwidth]{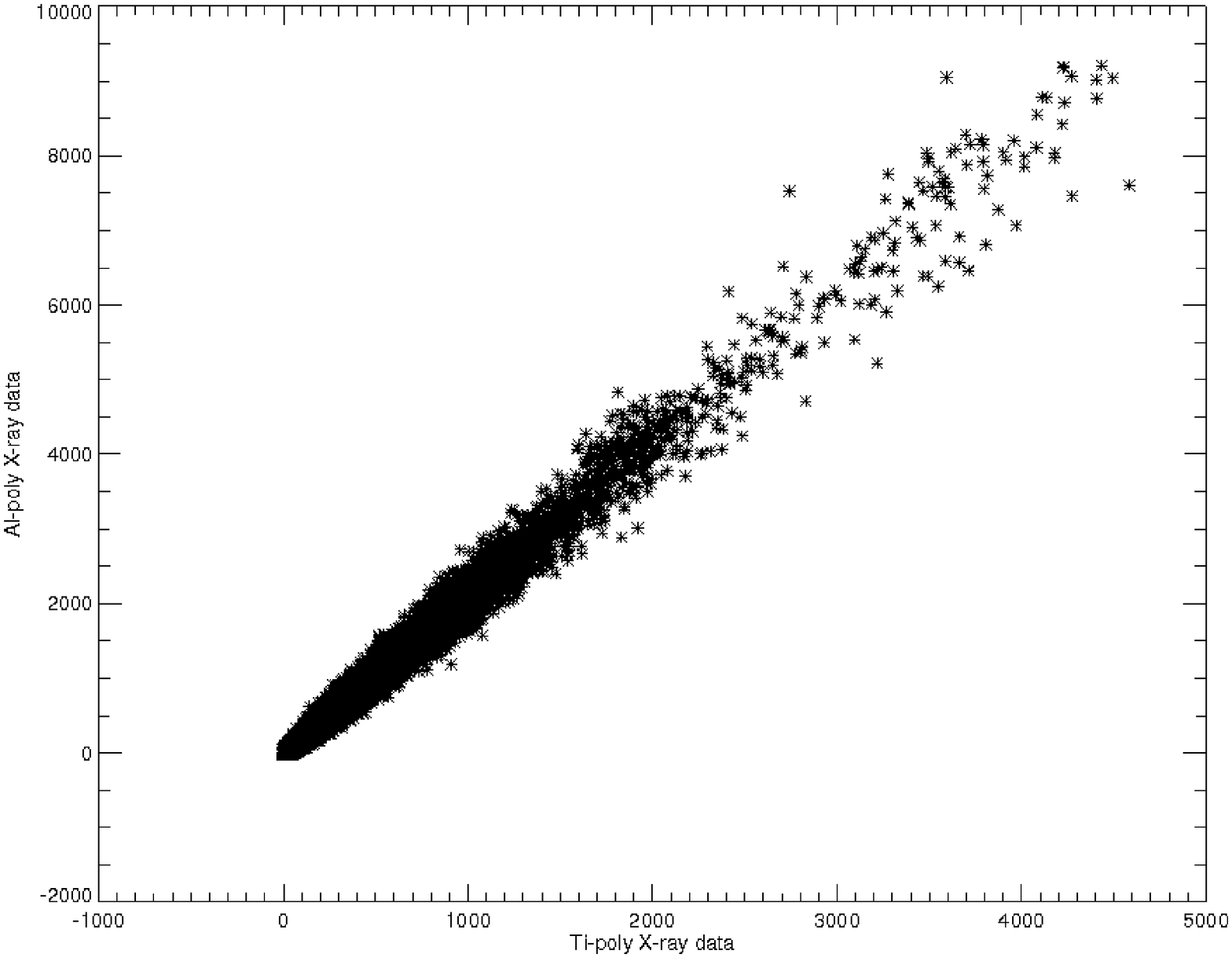}
\end{center}
\caption{\footnotesize X-ray data obtained from the Ti-poly and Al-poly filters. The linear correlation coefficient is 0.99.}
\label{fig:7}
\end{figure}

\section{Summary and Discussion}

A dominant fraction of coronal X-ray emission is known to originate within strongly magnetized active-region structures. To establish which of the magnetic-field quantities within these active regions contributes to the observed X-ray brightness, we have analysed the X-ray data from the XRT instrument and vector magnetic field measurements from the SP instrument onboard the {\it Hinode} spacecraft. We observed a good correlation between the total area-integrated magnetic field parameters and the
X-ray brightness. A strong correlation is observed  with the total unsigned magnetic flux and closer inspection indicates that other magnetic parameters are correlated with the X-ray brightness through their dependence on magnetic flux. This establishes that the magnetic flux (and thus size) of the system matters. It is generally observed that larger active regions have higher magnetic flux than smaller active regions, which suggests that larger active regions are brighter in X-rays than the small active regions. This result reconfirms the earlier result of \inlinecite{fish98}, which was based on lower resolution data and is valid across a range of orders of magnitudes across stars and other astrophysical objects \cite{pevt03}.

A large amount of total current is indicative of a highly non-potential active region with a large reservoir of energy. Does this larger energy reserve due to non-potentiality directly translate into stronger coronal X-ray emission?. Large-scale current systems are known to produce large-scale flares \cite{sch08}. It has previously been shown by \inlinecite{nand03} that the variance in the distribution of the local twist within active-region flux systems is also an indicator of the flare productivity of active regions. However, even if this background is suggestive of the role of active region non-potentiality in the release of energy and one might surmise also in coronal heating, we did not find this to be the case here. In fact, we found a (weak) negative correlation between X-ray flux and a measure of non-potentiality, namely $\mu_0 J_{\textrm{tot}}/\phi_{\textrm{tot}}$. If it does really exist, this correlation has no obvious explanation (at least at this time). Previous studies have shown that active-region non-potentiality has a stronger correlation with flare productivity than magnetic flux \cite{song06, jing06}. On the other hand, we found a stronger correlation between X-ray brightness and unsigned magnetic flux. Thus, one can argue that while non-potentiality may be an important determinant of localized heating related to flare productivity, the total unsigned magnetic-flux content is the primary factor governing large-scale coronal heating over the active regions.

For the Alfv\'en wave-heating model ({\it e.g.} see the review by \opencite{ascn04}), magnetic flux is related to the power dissipated at the active region through the square of the Alfv\'en velocity, whereas the X-ray brightness would be some fraction of this power -- which also indicates that there should be a relationship between total X-ray brightness and the total magnetic flux. However, based on a detailed analysis, \inlinecite{fish98} showed that the energy in these waves is not sufficient to explain the observed level of coronal heating. The MCC model \cite{long96} also predicts a strong correlation between total X-ray brightness and total magnetic flux. On the other hand, in the nano-flare heating  model \cite{park88} the power dissipated in active-region coronae is related to $B_{z,\textrm{tot}}^2$, suggesting that the total X-ray brightness would be strongly correlated with $B_{z,\textrm{tot}}^2$ rather than $\phi_{\textrm{tot}}$.

Our observations and analysis suggest that the MCC model is a viable contender as a physical theory for the heating of solar and stellar coronae. Nevertheless, we note that it is very likely that a variety of physical processes may contribute to coronal heating to different extents; there are numerous other subtleties in the coronal-heating problem that are far from being settled and need further investigations.

\section*{Acknowledgements}

{\it Hinode} is a Japanese mission developed and launched by ISAS/JAXA, with NAOJ as a domestic partner and NASA and STFC (UK) as international partners. It is operated by these agencies in cooperation with ESA and the NSC (Norway). We are also grateful to Bruce Lites and the CSAC team for the {\it Hinode}/SP data. The Center of Excellence in Space Science India (CESSI) is supported by the Ministry of Human Resource Development, Government of India. SH is grateful to the Council for Scientific and Industrial Research, University Grants Commission, Government of India for financial support. DN acknowledges a Ramanujan Fellowship from the Department of Science and Technology and the (US) Asian Office of Aerospace Research and Development for financial support. Finally, we thank an anonymous referee, David E. McKenzie, Edward DeLuca, and the XRT team for critical comments and many useful discussions.


\begin{thebibliography}{}

\bibitem[\protect\citeauthoryear{Aschwanden} {2004 }]{ascn04} Aschwanden, M.: 2004, {\it Physics of the Solar Corona}, An Introduction, Springer, Praxis Publishing, Chichester, UK.
\bibitem[\protect\citeauthoryear{Chandrashekhar {\it et al.}} {2013}]{chand13} Chandrashekhar, K., Krishna Prasad, S., Banerjee, D., Ravindra, B., Seaton, D.B.: 2013, {\it Solar Phys.} {\bf 286}, 125. doi:10.1007/s11207-012-0046-1 
\bibitem[\protect\citeauthoryear{Cirtain {\it et al.}} {2013}]{cirt13}
Cirtain, J.W., Golub, L., Winebarger, A.R., De Pontieu, B., Kobayashi, K., Moore, R.L. {\it et al.}: 2013, {\it Nature} {\bf 493}, 501. doi:10.1038/nature11772
\bibitem[\protect\citeauthoryear{Falconer et al.} {1997}]{falc97}
Falconer, D.A., Moore, R.L., Porter, J.G., Gary, G.A., Shimizu, T.: 1997, {\it Astrophys. J.} {\bf 482}, 519. doi:10.1086/304114 
\bibitem[\protect\citeauthoryear{Falconer} {1997}]{falc97a}
Falconer, D.A.: 1997, {\it Solar Phys.} {\bf 176}, 123. doi:10.1023/A:1004989113714
\bibitem[\protect\citeauthoryear{Falconer {\it et al.}} {2000}]{falc20}
Falconer, D.A., Gary, G.A., Moore, R.L., Porter J. G.: 2000, {\it Astrophys. J.} {\bf 528}, 1004. doi:10.1086/308188
\bibitem[\protect\citeauthoryear{Fisher {\it et al.}} {1998}]{fish98}
Fisher, G.H., Longcope, D.W., Metcalf, T.R., Pevtsov, A.A.: 1998, {\it Astrophys. J.} {\bf 508}, 885. doi:10.1086/306435
\bibitem[\protect\citeauthoryear{Golub {\it et al.}} {2007}]{golu07}
Golub, L., DeLuca, E., Austin, G., Bookbinder, J., Caldwell, D., Cheimets, P. {\it et al.}: 2007, {\it Solar Phys.} {\bf 243}, 63. doi:10.1007/s11207-007-0182-1
\bibitem[\protect\citeauthoryear{Hagyard} {1988}]{hagya88}
Hagyard, M.J.: 1988, {\it Solar Phys.} {\bf 115}, 107. doi:10.1007/BF00146233
\bibitem[\protect\citeauthoryear{Ichimoto {\it et al.}} {2008}]{ichim08}
Ichimoto, K., Lites, B., Elmore, D., Suematsu, Y., Tsuneta, S., Katsukawa, Y. {\it et al.}:  2008, {\it Solar Phys.} {\bf 249}, 233. doi:10.1007/s11207-008-9169-9
\bibitem[\protect\citeauthoryear{Jing {\it et al.}} {2006}]{jing06}
Jing, J., Song, H., Abramenko, V., Tan, C., Wang, H.: 2006, {\it Astrophys. J.} {\bf 644}, 1273. doi:10.1086/503895
\bibitem[\protect\citeauthoryear{Klimchuk} {2006}]{klim06} Klimchuk, J.A.: 2006, {\it Solar Phys.} {\bf 234}, 41. doi:10.1007/s11207-006-0055-z
\bibitem[\protect\citeauthoryear{Kosugi {\it et al.}} {2007}]{kosu07}
Kosugi, T., Matsuzaki, K., Sakao, T., Shimizu, T., Sone, Y., Tachikawa, S. {\it et al.}: 2007, {\it Solar Phys.} {\bf 243}, 3. doi:10.1007/s11207-007-9014-6
\bibitem[\protect\citeauthoryear{Lee {\it et al.}} {2010}]{lee10}
Lee, J.-Y., Barnes, G., Leka, K.D., Reeves, K.K., Korreck, K.E., Golub, L. {\it et al.}: 2010, {\it Astrophys. J.} {\bf 723}, 1493. doi:10.1088/0004-637X/723/2/1493
\bibitem[\protect\citeauthoryear{Leka, Fan, and Barnes} {2005}]{leka05} Leka, K.D., Fan, Y., Barnes, G.: 2005, {\it Astrophys. J.} {\bf 626}, 1091. doi:10.1086/430203
\bibitem[\protect\citeauthoryear{Leka, Barnes, and Crouch} {2009}]{Leka2009} Leka, K.D., Barnes, G., Crouch, A.: 2009, In: Lites, B., Cheung, M., Magara, T., Reeves, K. (eds.) \textit{The Second Hinode Science Meeting: Beyond Discovery-Toward Understanding}. {\bf CS-415}, Astron. Soc. Pacific, San Francisco, 365.

\bibitem[\protect\citeauthoryear{Leka and Barnes} {2007}]{leka07}Leka, K.D., Barnes, G.D.: 2007, {\it Astrophys. J.} {\bf 656}, 1173. doi:10.1086/510282
\bibitem[\protect\citeauthoryear{Lin, Kuhn, and Coulter} {2004}]{lin04} Lin, H., Kuhn, J.R., Coulter, R.: 2004, {\it Astrophys. J. Lett.} {\bf 613}, L177. doi:10.1086/425217
\bibitem[\protect\citeauthoryear{Longcope} {1996}]{long96}
 Longcope, D.W.: 1996, {\it Solar Phys.} {\bf 169}, 91. doi:10.1007/BF00153836
\bibitem[\protect\citeauthoryear{Longcope, Fisher, and Pevtsov} {1998}]{long98}
 Longcope, D.W., Fisher, G.H., Pevtsov, A.A.: 1998, {\it Astrophys. J.} {\bf 507}, 871. doi:10.1086/306312
\bibitem[\protect\citeauthoryear{Lundquist {\it et al.}} {2008}]{lund08}
 Lundquist, L.L., Fisher, G.H., Metcalf, T.R., Leka, K.D., McTiernan, J.M.: 2008, {\it Astrophys. J.} {\bf 689}, 1388. doi:10.1086/592760
\bibitem[\protect\citeauthoryear{Mandrini, Demoulin and Klimchuk} {2000}]{mandr20}
Mandrini, C.H., Demoulin, P., Klimchuk, J.A.: 2000, {\it Astrophys. J.} {\bf 530}, 999. doi:10.1086/308398
\bibitem[\protect\citeauthoryear{McIntosh {\it et al.}} {2011}]{mcint11} 	
McIntosh, S.W., de Pontieu, B., Carlsson, M., Hansteen, V., Boerner, P., Goossens, M.: 2011, {\it Nature} {\bf 475}, 477. doi:10.1038/nature10235 
\bibitem[\protect\citeauthoryear{Metcalf} {1994}]{metc94}
Metcalf, T.R.: 1994, {\it Solar Phys.} {\bf 155}, 235. doi:10.1007/BF00680593
\bibitem[\protect\citeauthoryear{Metcalf {\it et al.}} {1994}]{metc94b}
Metcalf, T.R., Canifield, R.C., Hudson, H.H., Mickey, D.L., Wulser, J.-P., Martens, P.C.H {\it et al.}: 1994, {\it Astrophys. J.} {\bf 428}, 860. doi:10.1086/174295
\bibitem[\protect\citeauthoryear{Nandy} {2008}]{nandy08} Nandy, D.: 2008,
In: Howe, R. Komm, R.W., Balasubramaniam K.S., Petrie, G.J.D. (eds.)
{\it Subsurface and Atmospheric Influences on Solar Activity} {\bf CS-383}, Astron. Soc. Pacific, San Francisco, 201.
\bibitem[\protect\citeauthoryear{Nandy \it{et al.}} {2003}]{nand03} Nandy, D., Hahn, M., Canifield, R.C., Longcope, D.W.: 2003, {\it Astrophys. J. Lett.} {\bf 597}, L73.  doi:10.1086/379815
\bibitem[\protect\citeauthoryear{Narain and Ulmscheider} {1996}]{nar96} Narain, U., Ulmscheider, P.: 1996, {\it Space Sci. Rev.} {\bf 75}, 453. doi:10.1007/BF00833341
\bibitem[\protect\citeauthoryear{Parker} {1988}]{park88}
Parker, E.N: 1988, {\it Astrophys. J.} {\bf 330}, 474. doi:10.1086/166485
\bibitem[\protect\citeauthoryear{Pallavicini {\it et al.}} {1979}]{palla79}
Pallavicini, R., Vaiana, G.S., Tofani, G., Felli, M.: 1979, {\it Astrophys. J.} {\bf 229}, 375. doi:10.1086/156963
\bibitem[\protect\citeauthoryear{Pevtsov {\it et al.}}{2003}]{pevt03}
Pevtsov, A.A., Fisher, G.H., Acton, L.W., Longcope, D.W., Johns-Krull, C.M., Kankelborg, C.C. {\it et al.}: 2003, {\it Astrophys. J.} {\bf 598}, 387. doi:10.1086/378944
\bibitem[\protect\citeauthoryear{Schrijver {\it et al.}} {2006}]{sch06}
Schrijver, C.J., De Rosa, M.L., Metcalf, T.R., Liu, Y., McTiernan, J., R{\'e}gnier, S. {\it et al.}: 2006, {\it Solar Phys.} {\bf 235}, 161. doi:10.1007/s11207-006-0068-7
\bibitem[\protect\citeauthoryear{Schrijver {\it et al.}} {2008}]{sch08}
Schrijver, C.J., De Rosa, M.L., Metcalf, T., Barnes, G., Lites, B., Tarbell, T. {\it et al.}: 2008, {\it Astrophys. J.} {\bf 675}, 1637.  doi:10.1086/527413
\bibitem[\protect\citeauthoryear{Song {\it et al.}} {2006}]{song06}
Song, H., Jing, J., Tan, C., Wang, H.: 2006, {\it AAS/Solar Physics Division Meeting} {\bf 37}, 09.05
\bibitem[\protect\citeauthoryear{Tan {\it et al.}}{2007}]{tan07}
Tan, C., Jing, J., Abramenko, V.I., Pevtsov, A.A., Song, H., Park, S.-H. {\it et al.}: 2007, {\it Astrophys. J.} {\bf 665}, 1460. doi:10.1086/519304
\bibitem[\protect\citeauthoryear{Tomczyk {\it et al.}} {2007}]{tomc07} Tomczyk, S., McIntosh, S.W., Keil, S.L., Judge, P.G.,
Schad, T., Seeley, D.H., Edmondson, J.: 2007, {\it Science} {\bf 317}, 1192. doi:10.1126/science.1143304 
\bibitem[\protect\citeauthoryear{Tsuneta {\it et al.}} {2008}]{tsun08}
Tsuneta, S., Ichimoto, K., Katsukawa, Y., Nagata, S., Otsubo, M., Shimizu, T. {\it et al.}: 2008, {\it Solar Phys.} {\bf 249}, 167. doi:10.1007/s11207-008-9174-z
\bibitem[\protect\citeauthoryear{Vaiana {\it et al.}} {1973}]{vaia73}
Vaiana, G.S., Davis, J.M., Giacconi, R., Krieger, A.S., Silk, J.K., Timothy, A.F.: 1973, {\it Astrophys. J. Lett.} {\bf 185}, L47. doi:10.1086/181318
\bibitem[\protect\citeauthoryear{Venkatakrishnan and Gary} {1989}]{venkat89} Venkatakrishnan, P., Gary, G.A.: 1989, {\it Solar Phys.} {\bf 120}, 235. doi:10.1007/BF00159877
\bibitem[\protect\citeauthoryear{Wang {\it et al.}} {2000}]{wan20}
 Wang, H., Yan, Y., Sakurai, T., Zhang, M.: 2000, {\it Solar Phys.} {\bf 197}, 263. doi:10.1023/A:1026566710505
\bibitem[\protect\citeauthoryear{Wang {\it et al.}} {2008}]{wan08}
 Wang, H., Jing, J., Changyi, T., Wiegelmann, T., Kubo, M.: 2008, {\it Astrophys. J.} {\bf 687}, 658. doi:10.1086/592082
\bibitem[\protect\citeauthoryear{Wedemeyer-B{\"o}hm et al.} {2012}]{wedem12} 
Wedemeyer-B{\"o}hm, S., Scullion, E., Steiner, O., van der Voort, L.R., de La Cruz Rodriguez, J., Fedun, V. {\it et al.}: 2012, {\it Nature} {\bf 486}, 505. doi:10.1038/nature11202
\bibitem[\protect\citeauthoryear{Withbroe and Noyes} {1977}]{withbr77}Withbroe, G.L., Noyes, R.W.: 1977, {\it Annu. Rev. Astron. Astrophys.} {\bf 15}, 363. doi:10.1146/annurev.aa.15.090177.002051
\bibitem[\protect\citeauthoryear{Zirker} {1993}]{zirk93} Zirker, J.B.: 1993, {\it Solar Phys.} {\bf 148}, 43. doi:10.1007/BF00675534

 \end{thebibliography}
\end{document}